Article

# Real-time, non-contact, cellular imaging and angiography of human cornea and limbus with common-path Full-field/SD OCT

Viacheslav Mazlin[1], Peng Xiao[1,2], Jules Scholler[1], Kristina Irsch[3,4], Kate Grieve[3,4], Mathias Fink[1] and A. Claude Boccara[1,*]

[1]*Langevin Institute, ESPCI PARIS, PSL Research University, CNRS, 1 Rue Jussieu, Paris, 75005, France*

[2]*State Key Laboratory of Ophthalmology, Zhongshan Ophthalmic Center, Sun Yat-sen University, Guangzhou 510060, China*

[3]*Vision Institute/CIC 1423, Sorbonne University, UMR_S 968 / INSERM, U968 / CNRS, UMR_7210, 17 Rue Moreau, Paris, 75012, France*

[4]*Quinze-Vingts National Eye Hospital, 28 Rue de Charenton, Paris, 75012, France*

*claude.boccara@espci.fr

## Abstract

In today's clinics, a cellular-resolution view of the cornea can be achieved only with an *in vivo* confocal microscope (IVCM) in *contact* with the eye. Here, we present a common-path Full-field/Spectral-domain OCT microscope (FF/SD OCT), which, for the first time, enables cell-detail imaging of the entire ocular surface (central and peripheral cornea, limbus, sclera, tear film) without contact and in real time. The device, that has been successfully tested in human subjects, is now ready for direct implementation in clinical research. Real-time performance is achieved through rapid axial eye tracking and simultaneous defocusing correction. Images, extracted from real-time videos, contain cells and nerves, which can be quantified over a millimetric field-of-view, beyond the capability of IVCM and conventional OCT. In the limbus, Palisades of Vogt, vessels and blood flow can be resolved with high contrast without contrast agent injection. The fast imaging speed of 275 frames/s (0.6 billion pixels/s) allowed direct monitoring of blood flow dynamics, enabling creation of high-resolution velocity maps for the first time. Tear flow velocity and evaporation time could be measured without fluorescein administration.

# Introduction

The cornea is the front part of the eye acting as a clear "window" into the world. In order to maintain optical clarity, it has a complex and highly-specialized micromorphology of fibers, cells and nerves. The tear film protecting the cornea on top and limbus with blood flow supply at the periphery also play essential roles in maintaining corneal health. A small malfunction in any part of this sophisticated system may lead to a broad range of potentially blinding (4$^{th}$ leading cause of blindness worldwide[1]) corneal disorders: degenerative (keratoconus), inherited or infectious (bacterial, viral and fungal keratitis). Taking into account that the largest corneal blindness burden falls on developing countries[2], cost-effective disease prevention through early diagnosis and treatment, and through public health programs is preferable over costly surgical interventions[3]. However, early and precise diagnosis is frequently complicated as various pathologies, requiring different therapies, may show the same symptoms on a macroscopic level[4]. Clinical OCT[5,6,7] with microscopic axial resolution proved to be useful for differentiating between many anterior eye conditions, nevertheless number of pathologies left unaddressed, because of unmet need for high lateral resolution in *en face* view. The first screening method to identify specific disease biomarkers with isotropic micrometer-level resolution was *in vivo* confocal microscopy (IVCM)[8]. Today IVCM is used in clinical practice as an important quantitative tool[9]. Nevertheless, its use is frequently avoided, primarily because of the requirement for direct physical contact with the patient's eye preceded by ocular anesthesia. This results in discomfort for the patient, increased risk of corneal damage and risk of infection. Moreover, IVCM provides a limited field of view (FOV), well below 0.5 mm, resulting in a long examination time (required to locate the spot of interest). Despite its drawbacks, IVCM, has remained the only high-resolution corneal screening tool available in clinics, with no alternatives up to this moment.

Recently, we developed *in vivo* Full-field Optical Coherence Tomography (FFOCT) and demonstrated its capability to capture images of the central human cornea with resolved nerves, cells and nuclei without touching the eye[10]. Moreover, this technology, originating from contact *ex vivo* FFOCT[11,12,13,14], acquires high-resolution *en face* images directly without scanning and without lateral motion artifacts contrary to instruments based on standard Spectral-domain OCT (SDOCT)[7,15]. This combination of benefits (non-contact operation, cell-resolution, *en face* optical sectioning without motion artifacts) was achieved thanks to our full-field interferometry approach. Nevertheless, the first *in vivo* FFOCT design captured an image only when the optical path lengths of the two interferometer arms were perfectly matched, which occurred only when the cornea happened to land in the perfect position. Indeed, given that the *in vivo* cornea is constantly moving, even during so-called steady fixation, such matching only occurs at rare random moments, prohibiting consistent real-time imaging and visualization of the whole breadth and depth of the cornea, necessary for clinical use of FFOCT.

Here we demonstrate a combined common-path FFOCT / SDOCT microscope (FF/SD OCT), which tracks the axial position of the eye and matches the optical arm lengths of FFOCT in real time to allow consistent imaging of the entire ocular surface (central, peripheral and limbal cornea, limbus, sclera, tear film). We demonstrate that real-time, millimeter-field videos of central and peripheral *in vivo* corneas consistently reveal cells and nerves, which can be quantified according to the existing medical protocols for IVCM. This makes FF/SD OCT ready for clinical research and translation into practice as a non-contact OCT-based alternative to IVCM with higher resolution than conventional OCT, which can detect micrometer changes in the entire ocular surface. Moreover, we show that our setup provides the additional possibility of monitoring tear film evolution, opening a door to quantitative and non-contact assessment of dry eye conditions. Beyond the cornea, the instrument visualizes the scleral and limbal regions, important to stem cell storage and regeneration. FF/SD OCT can also directly view blood flow with high contrast and without fluorescein injection at a high frame rate of 275 frames per second (fps), which paired with its high-resolution capabilities, allows individual red blood cells to be followed. Furthermore, we demonstrate a method to reveal high-resolution blood flow velocity and orientation maps, which may lead the way to new "localized" diagnostic methodologies of scleral inflammation and new approaches to monitor therapeutic effects locally.

## Results

**Tracking eye position with common-path FF/SD OCT**

The central idea in a common-path FF/SD OCT interferometer is to detect the axial position of the cornea in real-time with SDOCT and use this information to optically match the arms of the FFOCT interferometer by moving the reference arm, leading to consistent FFOCT imaging of a moving *in vivo* cornea (**Fig. 1, Supplementary Video 1**). SDOCT, coupled to the microscope objectives, displays the locations of the corneal surface and other reflecting surfaces in the XZ plane with high axial resolution (< 3.9 µm) by acquiring 2D *cross-sectional* 1.25 mm × 2.7 mm images of backscattered light intensity. 2D SDOCT images are formed through rapid 100 kHz A-line scanning with a galvanometer mirror. Combined SDOCT and FFOCT share the same optical paths in the arms of the FFOCT interferometer, which is apparent in the SDOCT image by appearance of a common-path FF/SD OCT peak in addition to the conventional peaks from the cornea and reference mirror of FFOCT (**Fig. 2a,b,c,d**). As the common-path and reference mirror peaks never overlap, we can simultaneously detect them and calculate actual positions of the cornea, imaging depth inside the cornea, optimal reference position, actual reference position and error for validating and improving the feedback loop (**Fig. 2e, Supplementary Video 2**). Every $8.2 \pm 0.5$ ms (mean ± standard deviation (s.d.)) information about the current corneal position is sent to the FFOCT system, where it is used to correct optical mismatch between the interferometer arms.

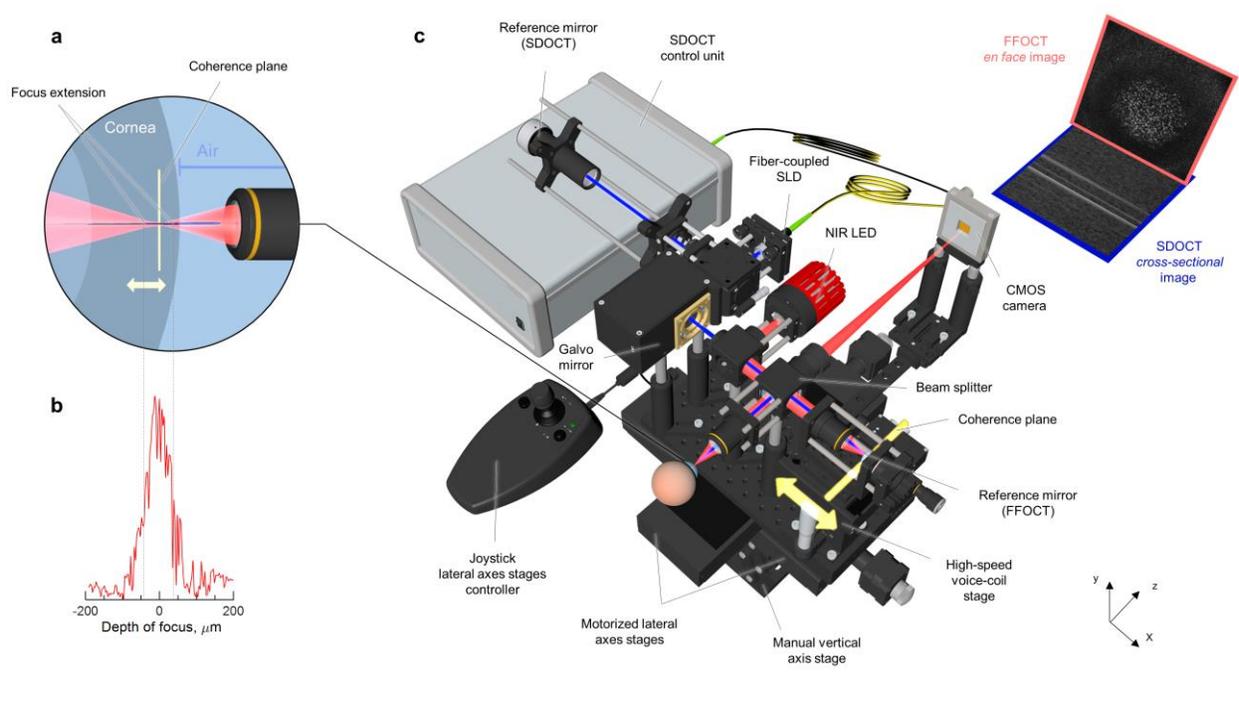

**Fig. 1 | Common path FF/SD OCT with axial eye tracking and real-time defocus adjustment. a**, Side-view of the FFOCT sample arm with the cornea. Location of the focus changes with changing corneal position, due to refraction at the air-cornea boundary (**Supplementary video 1**). The location of the coherence plane (depicted in yellow), corresponding to the position of the FFOCT reference arm, also shifts, but in the opposite direction, leading to an optical mismatch between the two and loss of FFOCT signal. **b**, Measured depth of focus. **c**, The microscope consists of two interferometers: FFOCT, which acquires *en face* images using a 850 nm light-emitting diode illumination (depicted in red), and SDOCT, capturing *cross-sectional* images with a 930 nm superluminescent diode light (depicted in blue). SDOCT data is used to calculate the current corneal position and the optical mismatch correction required, which is fed into the fast voice coil stage in the FFOCT reference arm. The stage shifts rapidly to place the coherence plane within the changing position of the depth of focus. As a result, the FFOCT interferometer arms match, and *en face* images are acquired consistently and in real-time (**Supplementary video 1**).

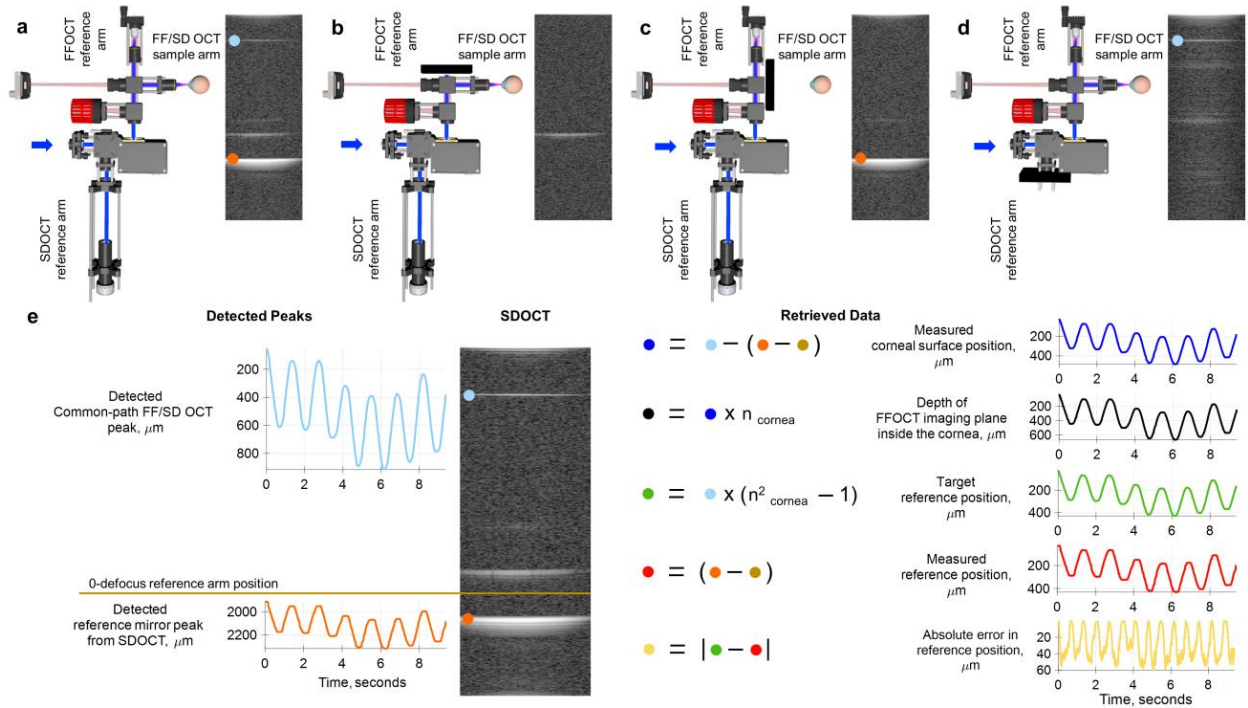

**Fig. 2 | Real-time detection of back-scattering intensity in SDOCT and calculation of corneal and reference mirror locations. a**, Left, common-path FF/SD OCT interferometer with light passing through all optical arms. **a** Right, SDOCT image with three planes of backscattered/reflected light. **b**, Left, the same device with blocked FFOCT reference arm. Right, SDOCT image with a single reflected peak, corresponding to the surface of the cornea. **c**, Left, device with blocked sample arm. Right, SDOCT image with reflection from the FFOCT reference mirror. **d**, Left, device with blocked SDOCT reference arm. Right, SDOCT image shows on top a single sharp reflection, originating from the FFOCT interference and captured by SDOCT (common-path peak). Its narrowness reflects the perfect dispersion matching in identical arms of FFOCT interferometer. **e**, Left, SDOCT image with common-path and reference mirror peaks being detected in real-time (**Supplementary video 2**). Both of these maxima move down when the reference arm is extended for defocusing correction. This facilitates their detection, as we can separate only the common-path peak for the upper area of the SDOCT image and the reference arm peak for the lower part without their overlapping. Right, using current locations of the two maxima and initial location of the reference mirror, one can calculate corneal surface position in real-time, along with the depth of the FFOCT imaging plane in the cornea, the optimal reference position for the current corneal location, the actual reference position and the error between the two, used for validating and improving the optical arms matching loop.

**Real-time optical path length matching of interferometer arms**

The FFOCT interferometer, equipped with a near-infrared (NIR) 850 nm incoherent light-emitting diode (LED) source and moderate numerical aperture (0.3 NA) 10× air microscope objectives (MO), acquires 2D *en face* 1.25 mm × 1.25 mm images of XY corneal sections with high lateral and axial resolutions (1.7 µm and 7.7 µm, respectively) by a time-domain two-phase shifting scheme[12] (**Fig. 1, Supplementary Video 1**). Each FFOCT image, composed of 1440 × 1440 pixels, is captured by a 2D CMOS camera in 3.5 ms at 275 FFOCT frames/s (or 0.6 billion pixels/s), which is 130 times faster (in terms of pixel rate) than the state-of-the-art corneal confocal scanning systems, imaging at 30 frames/s over 384 × 384 pixels FOV (or 3.6 million pixels/s)[16]. FFOCT alone does not detect the location of the cornea, and therefore this information is communicated from the SDOCT and is used to calculate the mismatch in the FFOCT interferometer arms (or more precisely, mismatch between the focus point of light in the sample and the coherence plane, located at the position of the optical path length, which matches with the reference arm length of the interferometer). The mismatch appears when the cornea is introduced into the sample arm and is a result of two factors: 1) the spreading of the focal point into the sample due to Snell's law, originating from the large illumination angle and difference between the refractive indexes of cornea (1.376) and air (1.0), 2) the shift of the coherence plane in the opposite direction closer to the objectives, also caused by the refractive index difference. From the known mismatch we calculate how far the reference arm of the interferometer should be extended, in accordance with the defocusing correction procedure[17], in order to match the coherence plane with a new position of the focus. This extension should be precise within the objective's depth of focus (measured ~ ± 35 µm at FWHM) (**Fig. 1b**) and rapid enough to follow *in vivo* movements of the cornea, which is achieved using a voice coil motor (accuracy = 2.2 µm, velocity = 1 mm/s, acceleration ~ 25 mm/s$^2$) operating at a 50 Hz update rate. The entire combined FF/SD OCT system is mounted on a manual vertical and two motorized horizontal stages, controllable by the operator using a joystick.

**Imaging of *ex vivo* cornea, mimicking the movements of the *in vivo* eye**

We tested and optimized real-time feedback of the common-path FF/SD OCT instrument by imaging *ex vivo* cornea (**Fig. 3i**), mounted on a moving stage. At first, the stage was programmed to produce a slow steady movement (20 µm/s) to compare system performance with defocus correction enabled and disabled. With defocusing correction, as the SDOCT detects the cornea coming closer to the objective, and the image plane going deeper into the sample, the voice coil stage extends the reference arm to put it in the optimal position for the current corneal location (**Supplementary video 3**). As a result, the FFOCT interferometer arms were well matched (**Fig. 3a**) with 3.7 ± 0.8 µm (mean ± s.d.) error well within the depth of focus, and FFOCT consistently displayed corneal images from various depths (**Fig. 3b**), while only occasionally the signal vanished due to phase changes induced by the movement of the sample. Conversely, without defocusing correction, the reference arm position remains fixed and FFOCT images are visible only at a single corneal position within the objective's depth of focus, which corresponds to matched optical path lengths of the arms of the interferometer (**Fig. 3c,d** and **Supplementary video 3**).

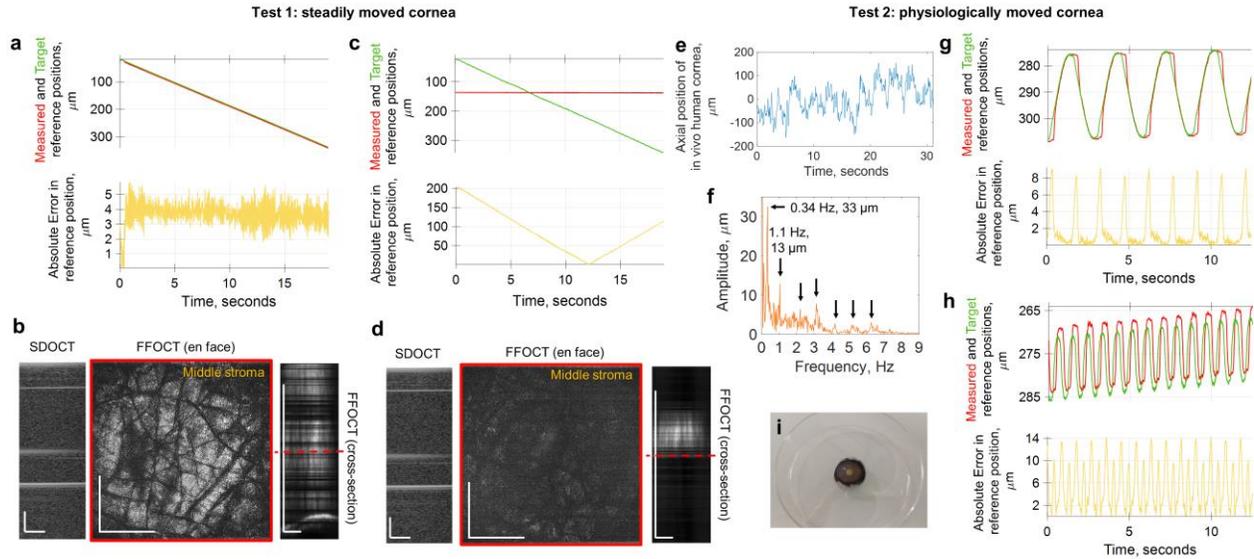

**Fig. 3 | Validating the real-time defocusing correction on *ex vivo* cornea, driven by the motorized stage. a**, Steadily moving *ex vivo* cornea, defocusing correction is active. Plots with measured reference mirror position (red), required reference location for ideal defocusing correction (green) and error between the two (yellow). **b**, Corresponding SDOCT, *en face* FFOCT and cross-sectional FFOCT images. Red line shows location of *en face* FFOCT image. FFOCT images are consistently acquired from various depths, while only occasionally the signal is vanishing due to additional phase introduced to the tomographic signal by the movement of the sample (**Supplementary video 3**). **c**, Plots for steadily moving *ex vivo* cornea, with defocusing correction being off. **d**, Without defocusing correction, the reference arm position was fixed and FFOCT images were visible only at a single corneal position within the objective's depth of focus, where the interferometric arms match. **e**, Axial position of the *in vivo* human cornea over time, measured with SDOCT. **f**, Extracted amplitudes and frequencies of the *in vivo* human corneal axial movements. **g**, Test of defocusing correction with *ex vivo* cornea physiologically moving at 0.34 Hz and with 33 µm amplitude. **h**, Test of defocusing correction with *ex vivo* cornea physiologically moving at 1.1 Hz and with 13 µm amplitude (**Supplementary video 4**). **i**, Photo of the corneal sample (without motor mount). Sample had visible stromal striae, indicative of tissue stress[18]. All scale bars 400 µm.

Next, we programmed the stage to move similarly to the physiological movements of the eye. This was achieved by first measuring the axial movements of the normal human cornea *in vivo* (**Fig. 3e**) and extracting the underlying frequencies and amplitudes (**Fig. 3f**). Two typical movements were visible: 1) heartbeat at 1.1 Hz with higher harmonics at 2.2 Hz, 3.3 Hz, etc., and 2) slow breathing at 0.34 Hz, both in agreement with the literature[19]. We used the frequencies with the two highest amplitudes (0.34 Hz with 33 µm amplitude and 1.1 Hz with 13 µm amplitude) as our input for the motorized stage with the sample (**Supplementary video 4**). In the two cases, defocusing correction demonstrated errors of 1.5 ± 2.2 µm (**Fig. 3g**) and 5.3 ± 4.0 µm (mean ± s.d.) (**Fig. 3h**), respectively, within the depth of focus. Higher frequencies were not important, as the corresponding amplitudes were smaller than the axial resolution as well as the depth of focus of the FFOCT device.

**Imaging of *in vivo* human cornea and sclera**

We applied common-path FF/SD OCT to view *in vivo* human cornea in real-time. The study was carried out on three healthy subjects (1 female and 2 males, aged 36, 24 and 26 years), which was confirmed by routine eye examination in the hospital preceding the experiment. Subjects expressed informed consent and the experimental procedures adhered to the tenets of the Declaration of Helsinki. Approval for the study was sought, in conformity to French regulations, from CPP (Comité de Protection de Personnes) Sud-Est III de Bron and ANSM (Agence Nationale de Sécurité du Médicament et des Produits de Santé) study number 2019-A00942-55. During the experiment, subjects were asked to rest their chin and temples on a standard headrest, while looking at a fixation target within the device. Examination was non-contact and without prior introduction of cycloplegic or mydriatic agents, nor topical anesthetics. Light irradiance was 86 mW/cm$^2$, below the maximum permissible levels of up-to-date ANSI and ISO ocular safety standards. Illumination was comfortable for viewing, due to the low sensitivity of the retina to NIR and IR light. The frame rate of real-time acquisition and display was ~ 10 frames/s, with each image captured in 3.5 ms. The operator of the instrument could simultaneously view FFOCT videos of *en face* images, OCT videos with *cross-sectional* images of the main corneal reflex, which is an indicator of the current imaging depth, and plots illustrating the performance of the defocusing correction. The instrument was able to acquire videos (**Supplementary videos 5,6,7,8**) from central, peripheral and limbal parts of the *in vivo* cornea (**Fig. 4a,o**), simultaneously correcting optical mismatch with 9.4 ± 6.2 µm, 11.3 ± 7.2 µm and 7.2 ± 6.6 µm (mean ± s.d.) errors (**Fig. 4b,u,v**), respectively.

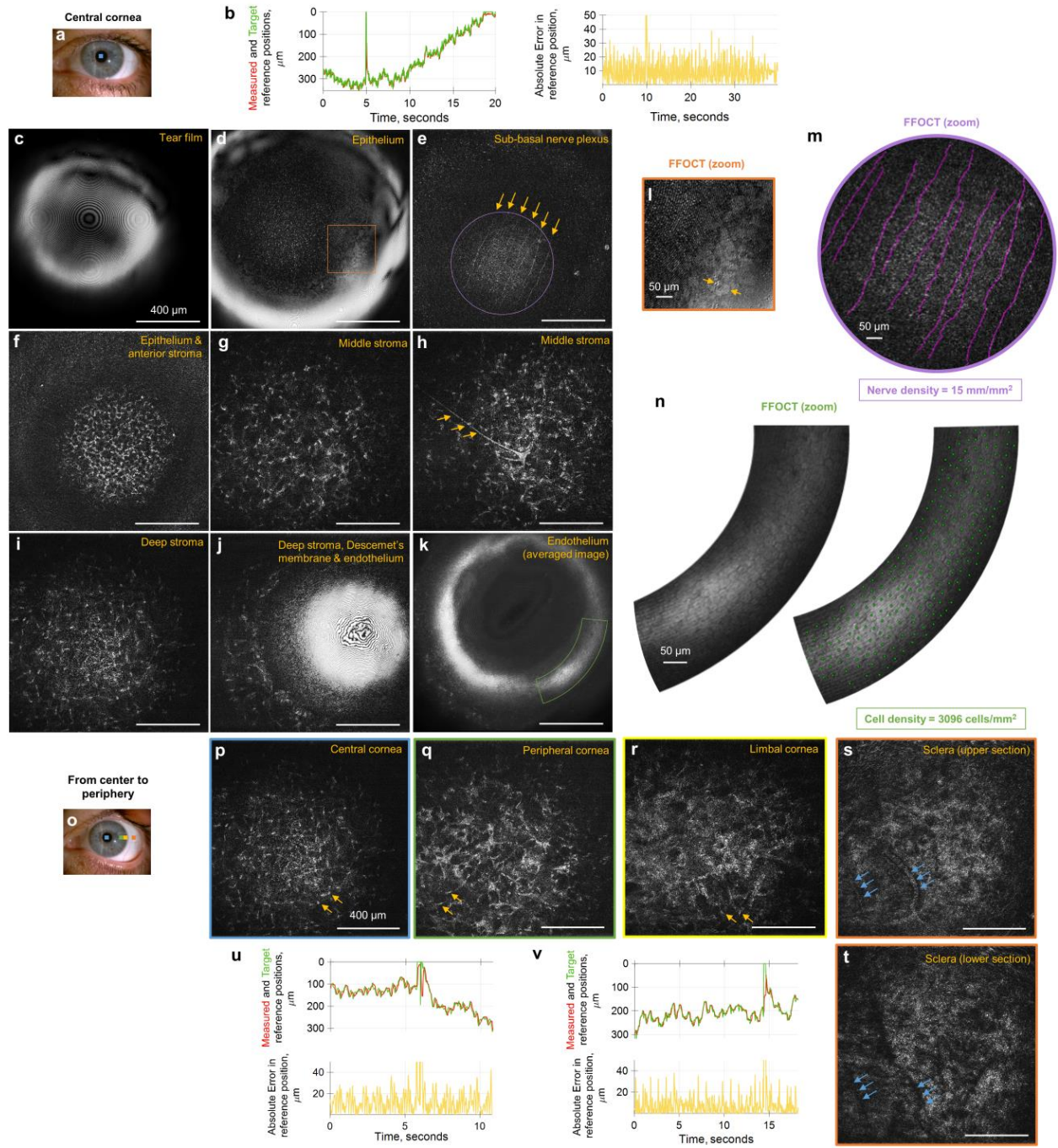

**Fig. 4 | Common-path FF/SD OCT imaging of central, peripheral human cornea and sclera *in vivo*. a**, Slit-lamp macro photo obtained from one of the subjects. Blue square depicts FFOCT field of view. **b**, Performance of real-time defocusing correction, when imaging central part of *in vivo* human cornea. Measured reference mirror position (red), required reference location for ideal defocusing correction (green) and error between the two (yellow). Peak at 10 seconds corresponds to the blink of the eye. **c-k**, Single frame FFOCT images through the entire thickness of the central cornea extracted from the real-time videos (**Supplementary videos 5,6**). Epithelial cells, sub-basal and stromal nerves (yellow arrows), keratocyte cells with nuclei and endothelial cells were visible. **l**, Zoomed and bandpass filtered FFOCT image of superficial epithelial cells with dark nuclei. **m**,

Zoomed FFOCT image of sub-basal nerves. NeuronJ[20] was used for nerve tracing and quantification. **n**, Zoomed FFOCT image of endothelial cell mosaic. ImageJ[21] point tool was used for cell counting. **o**, Slit-lamp macro photo with peripheral locations where FFOCT images were acquired. **p-t**, Single frame FFOCT images from central cornea to periphery, extracted from real-time videos (**Supplementary videos 7,8**). Resolving individual keratocyte nuclei (yellow arrows) was increasingly more difficult, when imaging further from the center. Blue arrows show vessels and their shadows in the sclera. **u, v**, Performance of real-time defocusing correction, when imaging the peripheral part of *in vivo* human cornea. Peaks at 6 and 14 seconds correspond to the blinks of the eye. All unlabeled scale bars 400 µm.

Non-averaged single FFOCT frames, extracted from the videos, had sufficiently high signal in all corneal layers. In the central cornea we observed tear film. It appears with a fringe pattern, as FFOCT is an interferometric technique, meaning that interference fringes are visible on flat surfaces such as tear film (**Fig. 4c**). Right below, we could see superficial epithelial cells 40 - 50 µm in diameter with dark 8 – 13 µm nuclei (**Fig. 4d,l**). These cells were revealed by image filtering in the Fourier domain (Online Methods). We could also see structures from other epithelial layers (wing, basal), however were unable to reliably resolve individual cells, due to low reflection contrast between them. Below the epithelium, we saw the sub-basal nerve plexus (SNP) with 2 – 4 µm thick nerves in a vertical orientation (**Fig. 4e**), which is characteristic of the central corneal area, located superior to the whorl-like nerve pattern[22]. Due to the curvature of the cornea and small thickness of the SNP, only part of the layer was visible. Nevertheless, the area of the visible nerve section was 0.317 mm$^2$, 3 times larger compared to what is possible with the state-of-the-art IVCM. As a result, a large-scale view of SNP can potentially be obtained with a smaller number of "stitched" images[23], leading to significantly faster screening and processing times. We also show that clinically valuable[24] measurement of nerve density can be performed with FFOCT. Nerves were segmented using NeuronJ[20] and their density measured 15 mm/mm$^2$ (**Fig. 4m**), within the healthy margins[25]. Underneath the SNP, we enter the anterior part of the stroma with numerous bright oval-shaped nuclei of keratocyte cells, measured about 15 µm in diameter (**Fig. 4f**). In the mid stroma, the density of keratocytes decreases and, in addition to the nuclei, we can resolve the cell bodies and branching 10 µm thick stromal nerves (**Fig. 4g,h**). With increasing depth, nuclei become more elongated and their density decreases (**Fig. 4i**). Descemet's membrane was also visible as a dark band separating stromal keratocytes from the endothelial cells (**Fig. 4j**). Endothelium viewed in a single FFOCT image was hindered by a strong specular reflection; nevertheless, on an averaged image we could resolve the hexagonal mosaic of 20 µm diameter cells and sometimes a 5 µm nucleus (**Fig. 4k**). Moreover, we were able to perform clinically significant[26] cell counting and measured the normal endothelial cell density of 3096 cells/mm$^2$ (**Fig. 4n**), in agreement with the literature[27]. All of the above images were acquired without any physical contact with the eye, with a distance of about 2 cm between the cornea and the microscope objective. We also benefited from the insensitivity of FFOCT to aberrations[28], more precisely the FFOCT resolution remained unchanged through the entire cornea, despite the presence of spherical and astigmatic aberrations.

We also looked at the appearance of stroma in central and peripheral cornea, and sclera. Initially the dark background of central corneal stroma (**Fig. 4p**), becomes bright at the periphery (**Fig. 4q,r**), which is explained by increased light scattering from the stromal fibrils, irregular in diameter and arrangement, known from electron microscopy studies[29,30]. Keratocyte cell nuclei, easily visible in central cornea, are more difficult to resolve in the periphery. Blood vessels (**Fig. 4s**), visible with their shadows (**Fig. 4t**), were perforating the upper sections of the sclera.

**Imaging of *in vivo* human tear film**

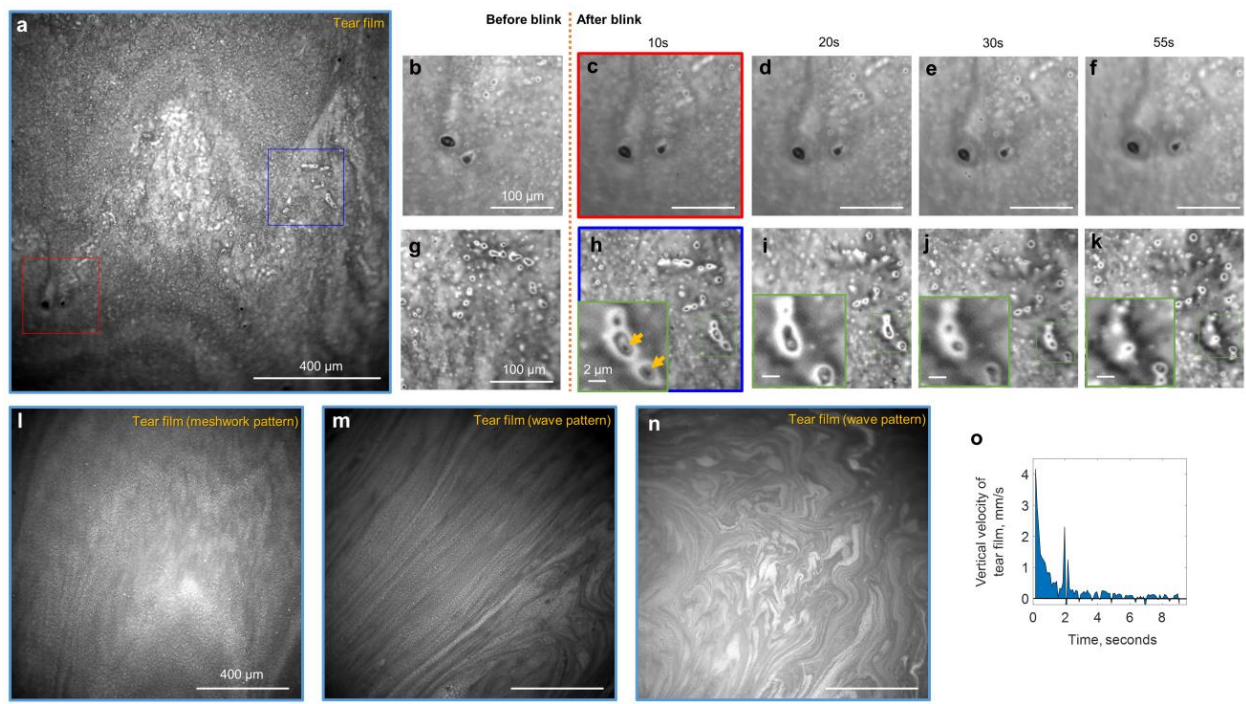

**Fig. 5 | Imaging human tear film *in vivo* with conventional microscope configuration of FF/SD OCT. a, l-n**, Single frame direct reflection images from the tear film. Interference patterns on the surface can be used to evaluate thickness of the lipid tear layer. **b-f, g-k**, Zoomed images from the real-time blink video (**Supplementary video 9**). Isolated particular matter, thought to be cellular debris or accumulations of newly secreted lipid from the Meibomian glands, were found in all the subjects. Particles were static, changing locations only from blink to blink together with the movement of the tear film. Green zoomed images show liquid drops surrounding small particles. Liquid was evaporating over time. **o**, Vertical velocity of the tear film, measured by manually tracking movements of particles in the video. Tear film stabilized after 9 seconds following the blink. Peak at 2 seconds corresponds to saccadic eye motion.

We were able to demonstrate tear film evolution by blocking the reference arm, thus converting our FFOCT setup into a conventional microscope. In order to increase contrast, an image of stray light from the beam splitter, acquired without the sample, was subtracted from the tear film image. Before each examination, subjects were asked to keep their eyes closed for 2 minutes to replenish the tear film. Right after the eye opened, an interference wave pattern was typically observed (**Fig. 5m,n**). In the condition of opened eyes with blinks present, we often saw meshwork (**Fig. 5a,l**) patterns. We acquired videos of tear flow after the blink (**Supplementary video 9**) and after the half-blink, when the eye was not completely closed, but the tear film shifted (**Supplementary video 10**). With the blink, the upper lid rapidly moved upward and the layer of tears followed with a slight delay. At 150 milliseconds after the blink, the flow velocity was 4.2 mm/s and rapidly decreased to 0.8 mm/s after 1 second, completely stabilizing to zero in 9 seconds, in agreement with literature[31,32] (**Fig. 5o**). Isolated particular matter in lipid and aqueous layers about 1 - 40 µm (**Fig. 5a,b,g**), thought to be cellular debris or accumulations of newly secreted lipid from the Meibomian glands[31], were found in all subjects. Particles were static (**Fig. 5c-f** and **Fig. 5h-k**), changing locations only from blink to blink together with the movement of the tear film (**Fig. 5b,c** and **Fig. 5g,h**). We also noticed that small particles were frequently surrounded by liquid drops, which were evaporating over time (**Fig. 5h-k**).

**Imaging of *in vivo* human limbus**

In the inferior limbal region, 30 µm wide radial palisades of Vogt (POV) were visible (**Fig. 6b,c**). The distance between the palisades measured 30 – 200 µm. Marginal corneal vascular arcades (MCA) with thin 3 - 7 µm vessels (**Fig. 6e**) and their dark shadows (**Fig. 6d**) appeared to protrude from inside of the palisades, in agreement with the literature[33,34,35]. These vessels, parallel to the ocular surface, appeared to be connected with a perpendicular oriented vessel network, visible as dark round shadows (**Fig. 6c,d**). Closer to the cornea, vessels were curling into the loops, while continuing to spread in the radial direction (**Fig. 6e**). Beneath, we could see thicker 40 µm branching vessels. By looking at the difference between two conventional images from the camera, it was possible to resolve individual 7 µm erythrocytes (**Fig. 6m**), which were difficult to see with an ordinary FFOCT image (**Fig. 6l**). Moreover, using rapid 275 frames/s acquisition, we could visualize and track the flow of these cells, which had a speed of about 1 mm/s (**Supplementary video 11**). Furthermore, in order to get the full-field information about the blood flow, important for addressing inflammatory conditions in the anterior eye, we measured the local cross-correlation for each 16 pixel × 16 pixel sub-image and retrieved blood flow velocity and blood flow direction maps with micrometer-level resolution (**Fig. 6h,i,j**). Average velocity was measured to be 0.446 ± 0.270 mm/s, with the lowest speeds close to vessel walls and highest speeds in the middle of the vessels, at junctions and also in locations where vessels were overlaying each other (**Fig. 6i**). Blood flow was mostly radial going towards or backwards from the cornea (**Fig. 6j**).

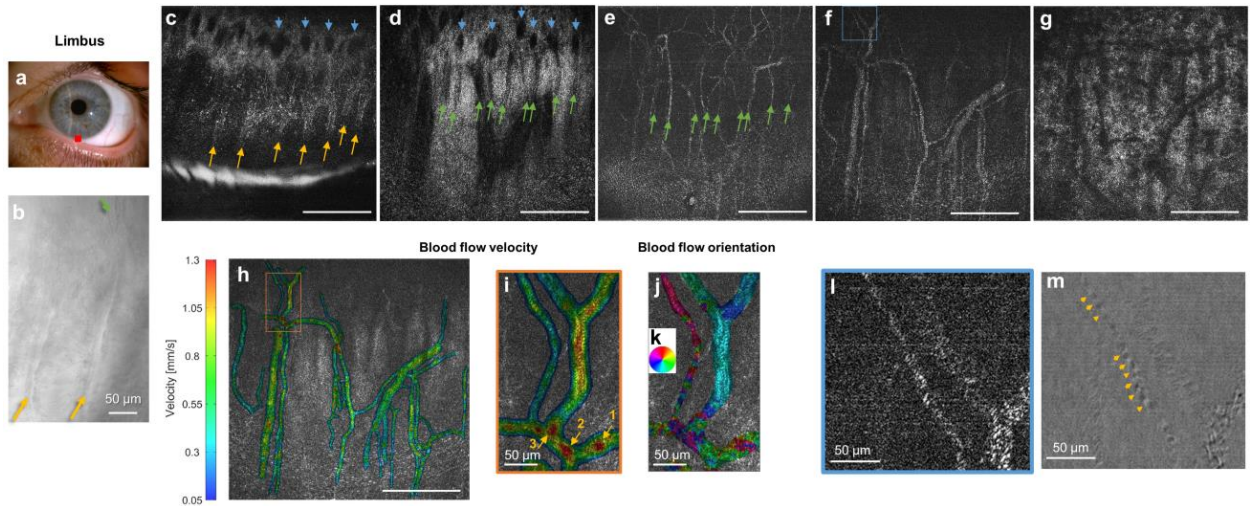

**Fig. 6 | Common-path FF/SD OCT imaging and angiography of *in vivo* human limbus. a**, Slit-lamp macro photo obtained from one of the subjects. Square depicts locations, where FFOCT images were acquired. **b**, Single frame direct reflection image of palisades of Vogt (yellow arrows) and vessels (green arrow) in limbus. **c-g**, Single frame FFOCT images of consecutively lower depths in limbus. Yellow arrows – palisades, green arrows – thin vessels hosted within palisades, blue arrows – perpendicular vessel network thought to be connected with horizontal vessels. Underneath, thicker vessels and scattering stroma-sclera medium with vessel shadows were seen. **h,i**, Blood flow velocity map, retrieved from the rapidly acquired video at 275 images/s (**Supplementary videos 11**), together with a zoomed image. The lowest speeds were measured close to vessel walls and highest speeds in the middle of the vessels, in junction points (point 2 and point 1 (with merging vessel coming to point 1 visible only in the video)) and also, as artifacts, if vessels were overlaying on top of each other (point 3). **j**, Blood flow orientation map retrieved from the video. Each color corresponds to a certain direction of blood flow, according to the colormap **k**. **l**, Zoomed FFOCT image of thin vessel in **f**. **m**, The same spot. Difference between the two direct reflection images (two conventional camera shots subtracted). Contrast is reduced, but is more intuitive for resolving individual erythrocytes (yellow arrows) (**Supplementary videos 11**). Unlabeled scale bars 400 μm.

## Discussion

Confocal microscopy, the state-of-the-art tool for cellular-resolution imaging of the cornea, requires contact with the eye, preceded by ocular anesthesia. Conventional OCT provides high axial resolution of corneal layers but does not resolve cells, and *en face* views suffer from alignment artefacts due to eye movements during scanning. Our common-path FF/SD OCT reveals the same micrometer corneal features (cell, nerves, nuclei), but in a completely non-contact way, with a 2 cm distance from the eye, and does not require use of any medication. This improves patient comfort, removes risk of corneal damage and risk of infection, and opens up the possibility for high-resolution imaging in a risk-sensitive population (eg young children, candidates for corneal transplant surgery with fragile corneas). Comparing to our previous preliminary work[10], in this article we have integrated an SDOCT into the FFOCT, which allowed to track the position of the eye and match the arms of the FFOCT interferometer in real-time. For the first time, this has enabled consistent real-time acquisition of corneal images and videos with sufficient signal and allowed exploration of the entire ocular surface (central, peripheral, and limbal cornea, as well as limbus and sclera). This opens a path for FFOCT implementation in clinical research and translation into practice. Our device provides images with a three times larger field of view than IVCM, which makes it easier to locate the clinical area of interest and follow the healing progress of the same location over time. Furthermore, a larger viewing area is useful for quantitative measurements of sub-basal nerve and endothelial cell densities, which were demonstrated with common-path FF/SD OCT following methodologies, used in confocal studies. Both are clinically valuable parameters: nerve density correlates with keratoconus, dry eye, several types of keratitis and diabetes[36], while endothelial cell density correlates with dystrophies, uveitis and acute ischemic stroke[37]. The instrument also opens a door to quantitative diagnosis of dry eye condition by providing information about tear film velocity and stabilization time following a blink, the evaporation time of the liquid micro-droplets on the surface of the eye, and potentially the thickness of the tear film by grading lipid interference patterns in a way, similar to Guillon Keeler Tear Film Grading System[38]. Common-path FF/SD OCT can resolve finer high-resolution sections of the limbus with distinguished layers of palisades of Vogt and underlying vessels, compared to UHR-OCT[35] and IVCM[39]. Furthermore, our instrument is not limited to diagnosis of "static" corneal disorders, but can potentially monitor inflammatory and scarring conditions affecting the dynamics of the blood flow in the eye. Previously, general information about the blood flow velocity and vessel dimensions from anterior segment fluorescein angiography (ASFA) and indocyanine green dye angiography (IGDA) proved to be useful for distinguishing between the various forms of scleral inflammation, in particular between severe episcleritis and diffuse anterior scleritis; or between peripheral corneal opacification and corneal thinning[40]. With FF/SD OCT, we can have access to the same information and can view the blood flow with high contrast, but without fluorescein injection and with the examination taking a fraction of a second. Furthermore, the ability of the instrument to capture high-resolution maps of blood flow velocities and orientations can potentially be useful for opening up new "localized" diagnostic methodologies and new ways to monitor effects of therapies locally. This could bring new insights about the anatomy and physiology of the vascular system in the anterior eye and, in particular about the limbal vessels and MCA supplying the cornea. Moreover, we suggest that the intravascular mapping of the blood flow with our device may be a promising platform for testing *in vivo* microfluidic theories[41]. Last, but not least, common-path FF/SD OCT can also be used for non-contact examination of various *in vivo* human, animal and *ex vivo* tissues.

## Online Methods

**FFOCT device**

The FFOCT device is based on an interference microscope in a Linnik configuration with identical microscope objectives in both arms of the interferometer. Objectives (LMPLN10XIR, Olympus, Japan) have a numerical aperture (NA) of 0.3, 10× magnification and are responsible for high lateral resolution of 1.7 µm (Rayleigh criterion at 850 nm), confirmed by measuring the visible diameters (FWHM) of imaged gold nanoparticles. The working distance of these objectives was 18 mm, sufficient to avoid the risk of accidental physical contact with the eye. Illumination is provided by an NIR 850 nm light-emitting diode (LED) source (M850LP1, Thorlabs, USA). The source has a 30 nm spectral bandwidth, resulting in an axial resolution of 7.7 µm in the cornea. Light from the LED is collected by an aspheric condenser lens (ACL12708U-B, Thorlabs, USA) and is focused on the back focal plane of the objective. Before entering the objective, light from the source is equally separated by the 50:50 beam splitter cube (BS) (BS014, Thorlabs, USA) into the sample and reference arms of the interferometer. The arms are slightly tilted from a perpendicular orientation in order to avoid specular reflection from the BS side. The objective in the reference arm focuses light onto an absorptive neutral density (ND) glass filter (NENIR550B, Thorlabs, USA), which plays the role of a single mirrored surface with 4% reflectivity. Use of an ND filter with OD 5.0 instead of a glass plate enables elimination of ghost reflections from the back surface of the filter. A low reflectivity value is chosen to achieve high detection sensitivity, which is maximized when the reflectivities of the mirror and of the sample match, in agreement with existing FFOCT signal to noise calculations[12]. Reflectivity of the cornea, estimated from the Fresnel relations, is around 2%. By using a reference mirror with a reflectivity of 4%, we can expect sensitivity close to the ideal condition in terms of the background illumination influence[12]. Light in the sample arm, backscattered from the different planes in the cornea, and light in the reference arm, backscattered from a single mirror plane, are collected by the objectives, and recombined on the BS. This results in interference, but only for the light coming from the corneal plane, and light coming from the reference mirror plane, which match in terms of optical path length. Temporal coherence length of the light source determines the thickness of interference fringe axial extension and concurrently optical sectioning precision, which in our case is 7.7 µm. Interfering and non-interfering light, arising from other planes of the cornea, are focused onto a sensor by the magnifying tube lens (AC254-250-B, Thorlabs, USA), leading to 14× overall system magnification. The sensor is a high-full well capacity (2Me-) 1440×1440 pixel CMOS camera (Q-2A750-CXP, Adimec, Netherlands), which captures each 2D image in 1.75 ms, eliminating the need for point-by-point scanning. In order to reach the expected axial sectioning in the system, interfering light should be selected from the background. To do so, we use a 2-phase shifting scheme, where we rapidly modulate the position of the reference mirror using a Piezo Mirror-shifter (STr-25/150/6, Piezomechanik GmbH, Germany), synchronized with the camera, to capture two consecutive images, which differ by the fringe pattern shifted by $\pi$, and subtract them. The absolute value of the resulting image contains only interfering light from a 7.7 µm thick section in the cornea. We can capture images in two modes: 1) fast acquisition mode for visualizing blood flow, where a sequence of 40 images is captured at 550 frames/s, and 2) slow mode for real-time imaging, where images are displayed live on screen at 10 frames/s. The LED illumination is not continuous, but pulsed in synchronization with camera acquisition in order to decrease eye exposure in accordance with ISO 15004-2: 2007 and ANSI Z80.36-2016 standards.

**SDOCT device**

The SDOCT device is based on a commercial general-purpose Spectral-Domain OCT (GAN510, Thorlabs, USA). It consists of an interferometer with a galvanometric mirror system (OCTP-900(/M), Thorlabs, USA), which rapidly scans a collimated light beam laterally at 100 kHz to form a 2D cross-sectional image. In order to increase the frame-rate, the lateral extension of the image was limited to 64 pixels. The 2D image is averaged over the lateral dimension and the resulting 1D data is processed with Labview Peak detector VI (National Instruments, USA) to locate the maxima. As these are single non-overlapping peaks, their positions can be detected with a precision better than 3.9 µm in the cornea, as determined by the Rayleigh criterion and spectral bandwidth of infrared (IR) 930 nm superluminescent diode (SLD). Because SDOCT interferometer and FFOCT interferometer share the optical arms, we detect not only the conventional SDOCT peaks corresponding to the reflection from the cornea and from the FFOCT reference mirror, but also the third common-path peak corresponding to the interference inside the FFOCT interferometer (between FFOCT sample and FFOCT reference arms). The latter maximum has the advantages that 1) it does not suffer from dispersion, due to the perfect symmetry between the arms of the FFOCT interferometer; 2) it is moving in the same direction on the 2D image as the reference arm, extended for the defocus correction. As a result, since the common-path and reference mirror peaks never overlap, we can simultaneously detect them and calculate positions of the cornea, the imaging depth inside the cornea, the optimal reference position for the defocus correction[17], the actual reference position and the error for validating and improving the feedback loop. By reducing the light entering into the reference arm of SDOCT, we suppress the conventional peak from the cornea to further facilitate acquisition of common-path and conventional reference arm peaks. Maxima were detected over a 1.25 mm lateral range, matching the FFOCT field of view and over a 2.7 mm axial range, determined by the SDOCT spectrometer. New information about the locations of the peaks was obtained every $8.2 \pm 0.5$ ms (mean ± standard deviation (s.d.)).

**Integrated FFOCT-SDOCT instrument**

SDOCT is optically integrated through the dichroic mirror (Edmund Optics, USA) into the illumination arm of the FFOCT. In order to block the SDOCT light from reaching the FFOCT camera, two filters with opposite spectral characteristics (low and high pass at 900 nm cutoff) (FELH0900 and FESH0900, Thorlabs, USA) are positioned at the entrance and exit of the FFOCT device. Glass windows (WG11050-B, Thorlabs, USA) are inserted into the reference arm of SDOCT to dispersion match with FFOCT sample and reference arms. The combined instrument is positioned on two high-load lateral translation stages (NRT150/M, Thorlabs, USA), controlled by a driver (BSC202, Thorlabs, USA) with a joystick (MJC001, Thorlabs, USA). Beneath, the laboratory jack (L490/M, Thorlabs, USA) is used to position the whole device vertically. The reference arm of the FFOCT device is mounted on a voice-coil translation stage (X-DMQ12P-DE52, Zaber, Canada), driven with 2.2 µm accuracy, 1 mm/s velocity and 25 mm/s$^2$ acceleration for rapid defocus adjustment. Information about the currently required defocus correction shift is communicated to the stage from the SDOCT system through the same personal computer (PC) to avoid time delays. The weak link in the communication is the limited read frequency of the stage encoder, which can accept 20 new positions/s – about 2 times slower compared to the rate of provided positions by SDOCT. FFOCT and SDOCT image acquisitions are controlled using two PCs, which are synchronized with $11 \pm 3$ ms (mean ± s.d.) precision through the NI-PSP protocol via the local network of the Langevin Institute. This enables simultaneous recording of FFOCT and SDOCT data.

**Software**

Custom code written in Labview 2014 and Thorlabs SpectralRadar SDK was used for FFOCT and SDOCT image acquisition and display, peak detection and motor control. MATLAB R2017a was used for plots and calculation of angiography maps. We utilized ImageJ 1.51p for image display and for measurements of cell and nerve densities.

**Artifact suppression and contrast enhancement**

In order to remove surface interference fringes and reveal superficial epithelial cells (**Fig. 4l**), we transformed the image into the Fourier domain and masked the bright spots, corresponding to fringes. Inverting this Fourier image with suppressed artifacts revealed the cells. The same artifacts were also visible at the outer endothelial layer. We were able to suppress them by averaging multiple (23) tomographic FFOCT images without Fourier domain conversion. Direct images of Palisades of Vogt (**Fig. 6l**) were obtained by first capturing a single conventional camera frame without the sample, and then subtracting this frame from the one with the sample. In this way, we increased the contrast of the image in post-processing through removing the light coming from the reference mirror. The final image (**Fig. 6m**) was obtained by subtracting two consecutive conventional images from the camera without taking their absolute value. In the resulting image, we lost half of the useful signal (pixels with negative values), but were able to resolve the individual blood cells.

**Quantitative image analysis**

Semi-automated nerve segmentation and density analysis (**Fig. 4m**) was performed with ImageJ[21] using the NeuronJ plugin[20]. Manual endothelial cell counting (**Fig. 4n**) was done with the Multi-point Tool in ImageJ. A Manual Tracking ImageJ plugin enabled manual blood cell segmentation, tracking and the subsequent cell velocity analysis (**Fig. 6m** and **Supplementary video 9**).

**Mapping blood flow velocity and orientation**

Previously, we demonstrated the possibility of local blood flow measurements from the conjunctival surface using manually drawn kymograph plots (i.e. plotting the vessel curvilinear abscissa against time)[42]. Unfortunately, each plot provided only a single local velocity value, so that obtaining a full-field velocity map would require considerable manual effort. Here we used a method based on a block-matching algorithm to track single features inside vessels, which allows rapid semi-automatic mapping of blood flow velocities and orientations in full-field. We cross-correlated 16 x 16 pixels windows and retrieved the cross-correlation maximum for each pixel in the vicinity corresponding to a maximal speed of 2 mm/s. Low cross-correlation peaks were discarded in order to remove artifacts and outliers. A velocity array is then created based on 8 frames, so that the velocity computed for each pixel is the average of the 7 velocities computed by the block-matching algorithm. Then the velocities and orientations are mapped on the hue channel, the FFOCT image is mapped on the value channel and the saturation is arbitrarily set to 0.8 for each pixel in order to construct the velocity and orientation map of the blood flow (**Fig. 6h-j**).

**Ex vivo cornea**

*Ex vivo* macaque cornea was obtained from the partner research institution as recuperated waste tissue from an unrelated experiment. Corneas were dissected from the ocular globes within two hours post-mortem and fixed in 2% paraformaldehyde prior to transfer to the imaging lab. Some edematous swelling occurred, causing enhanced visibility of stromal striae[18], indicative of tissue stress.

**Imaging in vivo**

Informed consent was obtained from all subjects and the experimental procedures adhered to the tenets of the Declaration of Helsinki. Examination was non-contact and no medication was introduced into the eye. Light illumination, visible as a dim red circular background, was comfortable for viewing, due to the low sensitivity of the retina to NIR and IR light. Corneal irradiance measured 86 mW/cm$^2$ below the maximal permissible exposures (MPEs) according to the up-to-date ocular safety standards: ISO 15004-2:2007 and ANSI Z80.36-2016. Focusing the light beam in the cornea results in a widely spread out low-irradiance beam on the retina, leading to an exposure level at 2% of the MPE limit. The subject was sitting in front of the system and was comfortably positioned with temple supports and a chin rest. While one eye was imaged, the second eye was fixating on a target. When imaging non-central parts of the cornea, the subject's head was tilted by the examiner to position the eye's surface plane perpendicular to the direction of the incoming light beam.

**Other instruments**

Macro images were obtained with a slit-lamp biomicroscope (Topcon, France Medical S.A.S.) using 10× magnification and the lowest illumination.

## Acknowledgements

This work was supported by the HELMHOLTZ synergy grant funded by the European Research Council (ERC) (610110), as well as from the European Union's Horizon 2020 research and innovation program under the Marie

Skłodowska-Curie grant agreement No 709104 (K.I.). We are grateful to Cristina Georgeon for assistance with acquiring the slit-lamp images. We thank Michel Paques and Jean-Marie Chassot for valuable discussions.

**Author Contributions**

V.M., M.F. and A.C.B. conceptualized the general idea. V.M., P.X. and A.C.B conceived and developed the optical design. V.M., J.S. and A.C.B. conceived the software and hardware architecture. V.M. wrote the software and performed the imaging experiments. K.I. and K.G. provided the corneal samples and guidance on usage and experiments. V.M., K.I., K.G. and A.C.B. analyzed the acquired data. J.S. conceived and developed algorithms for obtaining quantitative blood flow velocity and orientation maps. V.M. wrote the manuscript, and all the authors contributed with edits and revisions.

**Competing Interests statement**

The authors declare no competing financial interests.